\documentclass{tlp}

\usepackage{url}
\usepackage{amsmath}
\usepackage{graphicx}
\usepackage{multirow}
\usepackage{fancyvrb}
\usepackage{csquotes}
\usepackage{xcolor}
\usepackage{booktabs}
\newcommand{\ra}[1]{\renewcommand{\arraystretch}{#1}}

\newcommand{\partialfrac}[2]{\frac{\partial #1}{\partial #2}}

\begin{document}

\lefttitle{Schrijvers, van den Berg and Riguzzi}

\jnlPage{1}{8}
\jnlDoiYr{2021}
\doival{10.1017/xxxxx}

\title[Automatic Differentiation in Prolog]{Automatic Differentiation in Prolog\thanks{This project is partly funded by the Flemish Fund for Scientific Research (FWO).}}

\begin{authgrp}
\author{\sn{Schrijvers} \gn{Tom}}
\affiliation{KU Leuven}
\author{\sn{van den Berg} \gn{Birthe}}
\affiliation{KU Leuven}
\author{\sn{Riguzzi} \gn{Fabrizio}}
\affiliation{Universit\`a degli Studi di Ferrara}
\end{authgrp}


\maketitle

\begin{abstract}
Automatic differentiation (AD) is a range of algorithms to compute the numeric value
of a function's (partial) derivative, where the function is typically given as a
computer program or abstract syntax tree. AD has become immensely popular as part
of many learning algorithms, notably for neural networks.
This paper uses Prolog to systematically derive gradient-based forward- and
reverse-mode AD variants from a simple executable specification: evaluation of the
symbolic derivative. Along the way we demonstrate that several Prolog features
(DCGs, co-routines) contribute to the succinct formulation of the algorithm.
We also discuss two applications in probabilistic programming that are enabled by
our Prolog algorithms. The first is parameter learning for the Sum-Product Loop Language
and the second consists of both parameter learning and variational inference for probabilistic
logic programming.
\end{abstract}


\section{Introduction}

Kowalski's slogan ``algorithm = logic + control''
\citep{DBLP:journals/cacm/Kowalski79} has been an inspiration to express and
study algorithms in Prolog. Notable examples are the \emph{Logic Programming
Pearls} of \emph{Theory and Practice of Logic Programming}
(e.g., \cite{DBLP:journals/tplp/VandecasteeleJ03,DBLP:journals/tplp/Bruynooghe04,DBLP:journals/tplp/SchrijversF06}),
and more recently Prolog versions of SAT/SMT solving
\citep{DBLP:journals/tcs/HoweK12} and of backjumping
\citep{DBLP:journals/tplp/RobbinsKH21}. Perhaps one of the best known examples
and a direct inspiration of this work, is the elegant symbolic differentiation
approach of \citet{ClocksinMellish03} that originally appeared in 1981.

Following these footsteps, we present a Prolog version of
\emph{automatic differentiation} (AD), a range of
algorithms to compute the numeric value of a function's (partial) derivative,
where the function is typically given as a computer program or abstract syntax
tree. AD differs from numeric differentiation approaches, which are
approximative, and from symbolic differentiation, which yields a symbolic result:
AD computes the derivate value exactly, evaluating it at a specific point.
While originally conceived in 1964~\citep{Wengert64}, AD has become
immensely popular in recent years under the name of \emph{backpropagation}, the
algorithm at the heart of neural networks. It also has applications in logic
programming, for instance for learning parameters of probabilistic logic
programs.

This work is inspired by the approach of \cite{DBLP:journals/corr/abs-2212-11088},
which derives forward-mode- and reverse-mode AD from symbolic
differentiation using algebraic abstractions in the purely functional setting of Haskell.
We recast AD in a logic programming setting:
we proceed step by step from symbolic differentiation and show how to
incorporate different optimizations and obtain several forward-mode and
reverse-mode variants (Sections~\ref{sec:symbolic}--\ref{sec:destructive}).
Along the way we demonstrate that several Prolog features such as
definite clause grammars (DCGs) and co-routines
contribute to the succinct formulation of the algorithm. Co-routines in
particular make it easy to express reverse-mode variants in a single phase
where conventional presentations require two phases. While we use a minimal
expression language throughout the main developments to focus on the essence,
Section~\ref{sec:extensions} explains how additional primitives (e.g., $\sin$,
$\cos$) can be incorporated.
We also present two case studies of AD where our Prolog implementations\footnote{\url{https://github.com/birthevdb/ad-prolog-code}} can be used.
The first (Section~\ref{sec:spll}) implements parameter learning
for the recently proposed Sum-Product Loop Language (SPLL). The second (Section~\ref{sec:plp})
concerns parameter learning \emph{and} variational inference for probabilistic logic programming.
Finally, Section~\ref{sec:related} discusses related work and Section~\ref{sec:conclusion}
concludes.

\section{Symbolic Expressions and Their Evaluation}\label{sec:symbolic}

Throughout most of the paper, we use a minimal
expression grammar consisting of
literals \texttt{lit(N)}, with \texttt{N} a number,
variables \texttt{var(X)}, with \texttt{X} a variable identifier,
addition \texttt{add(E$_1$,E$_2$)}, and
multiplication \texttt{mul(E$_1$,E$_2$)}.

The \texttt{eval(E,Env,N)} predicate evaluates expressions \texttt{E}
to a numeric result \texttt{N} given an environment \texttt{Env} that maps the variables
to their value.
\begin{Verbatim}[frame=single,fontsize=\small]
eval(lit(N),    _Env,N) .
eval(var(X),    Env ,N) :- lookup(X,Env,N).
eval(add(E1,E2),Env ,N) :- eval(E1,Env,N1), eval(E2,Env,N2), N is N1 + N2.
eval(mul(E1,E2),Env ,N) :- eval(E1,Env,N1), eval(E2,Env,N2), N is N1 * N2.
\end{Verbatim}

To allow a $\mathcal{O}(1)$ lookup in the environment, we use natural numbers in
the range $[1,n]$ for the variable identifiers and represent the environment itself
as a term \texttt{env(N$_1$,\ldots,N$_n$)} whose arguments are the values of
the corresponding variables. Then, \texttt{lookup/3} can be defined as follows.
\begin{Verbatim}[frame=single,fontsize=\small]
lookup(X,Env,N) :- arg(X,Env,N).
\end{Verbatim}

\section{Symbolic Differentiation}

\noindent
\begin{minipage}{.5\textwidth}
As a warm-up we start with the symbolic differentiation of our expressions,
following the example of \citet[\S 7.11]{ClocksinMellish03} and the textbook
differentiation rules.
The predicate \texttt{symb(E,X,DE)} captures the symbolic
differentiation relation $\partialfrac{E}{X} =$ \texttt{DE}
and computes it in a naive way.
\end{minipage} %
\begin{minipage}{.5\textwidth}
\small
\begin{eqnarray*}
\partialfrac{n}{x} & = & 0 \\
\partialfrac{y}{x} & = & \left\{\begin{array}{ll} 1 & (x = y) \\ 0 & (x \neq y) \end{array}\right. \\
\partialfrac{e_1 + e_2}{x} & = & \partialfrac{e_1}{x} + \partialfrac{e_2}{x} \\
\partialfrac{e_1 \times e_2}{x} & = & e_2 \times \partialfrac{e_1}{x} + e_1 \times \partialfrac{e_2}{x} \\
\end{eqnarray*}
\end{minipage}

\begin{Verbatim}[frame=single,fontsize=\small]
symb(lit(_N),    _X, lit(0)).
symb(var(Y),     X,  lit(DE)) :-
     ( X == Y -> DE = 1 ; DE = 0).
symb(add(E1,E2), X,  add(DE1,DE2)) :-
     symb(E1,X,DE1), symb(E2,X,DE2).
symb(mul(E1,E2), X,  add(mul(E2,DE1),mul(E1,DE2)) :-
     symb(E1,X,DE1), symb(E2,X,DE2).
\end{Verbatim}

In order to prepare for actual AD algorithms, we refactor the above code
from \texttt{symb/3} to \texttt{symb/4}. The latter not only
returns the symbolic derivative but also reconstructs the original expression;
this structure becomes advantageous in Section~\ref{sec:fwdad}'s non-symbolic setting.
The combination of the original expression (\emph{primal}) and the (symbolic)
partial derivative (\emph{tangent}) is also known as a (here: symbolic)
\emph{dual number} in the AD literature.

\noindent
\begin{minipage}{.7\textwidth}
\begin{Verbatim}[frame=single,fontsize=\small]
symb(lit(N),    _X,lit(N),   lit(0)).
symb(var(Y),    X, var(Y),   lit(DF)) :-
  ( X == Y -> DF = 1 ; DF = 0).
symb(add(E1,E2),X,add(F1,F2),add(DF1,DF2)) :-
  symb(E1,X,F1,DF1), symb(E2,X,F2,DF2).
symb(mul(E1,E2),X,mul(F1,F2),DF) :-
  symb(E1,X,F1,DF1), symb(E2,X,F2,DF2),
  DF = add(mul(F2,DF1),mul(F1,DF2)).
\end{Verbatim}
\end{minipage}%
\begin{minipage}{.3\textwidth}
\begin{gather*}
\texttt{symb(E,X,F,DF)}
\\ \Leftrightarrow \\
\left\{\begin{array}{l} E = F \\ \partialfrac{E}{X} = DF \end{array}\right.
\end{gather*}
\end{minipage}

\section{Forward-Mode Automatic Differentiation}\label{sec:fwdad}

Many practical applications do not require the symbolic derivative, but its numeric
value at a point. Naively, this can be obtained by evaluation after symbolic derivation.
\begin{Verbatim}[frame=single,fontsize=\small]
ad_spec(E,X,Env,N,DN) :-
  symb(E,X,F,DF),
  eval(F,Env,N),
  eval(DF,Env,DN).
\end{Verbatim}
The idea of AD is to perform the two steps, symbolic derivation and numeric
evaluation, simultaneously, in order to avoid the intermediate symbolic result.
This is accomplished by \texttt{fwdad(E,X,Env,F,DF)}, where \texttt{F} and \texttt{DF}
represent a dual number with the original expression and partial derivative,
respectively,
evaluated with respect to the environment \texttt{Env}.
This definition is essentially
obtained from \texttt{ad\_spec/4} by unfold/fold transformations~\citep{10.1145/53990.54020}
(\ref{app:fwdad}).

\noindent
\begin{tabular}{l@{\hspace{0.02\textwidth}}l}
\begin{minipage}{.49\textwidth}
\begin{Verbatim}[frame=single,fontsize=\small]
fwdad(lit(N),_X,_Env,N,0).
fwdad(var(Y),X,Env,F,DF) :-
  lookup(Y,Env,F),
  ( X == Y -> DF = 1 ; DF = 0).
fwdad(add(E1,E2),X,Env,F,DF) :-
  fwdad(E1,X,Env,F1,DF1),
  fwdad(E2,X,Env,F2,DF2),
  F is F1 + F2,
  DF is DF1 + DF2.
fwdad(mul(E1,E2),X,Env,F,DF) :-
  fwdad(E1,X,Env,F1,DF1),
  fwdad(E2,X,Env,F2,DF2),
  F is F1 * F2,
  DF is F2 * DF1 + F1 * DF2.
\end{Verbatim}
\end{minipage}
&
\begin{minipage}{.49\textwidth}
Had we started from \texttt{symb/3} rather than 
\texttt{symb/4}, we'd have instead ended up with the following clause for 
\texttt{mul/2}.
\begin{Verbatim}[fontsize=\small]
    fwdad(mul(E1,E2),X,Env,DF) :-
      fwdad(E1,X,Env,DF1),
      fwdad(E2,X,Env,DF2),
      eval(E1,Env,F1),
      eval(E2,Env,F2),
      DF is F2 * DF1 + F1 * DF2.
\end{Verbatim}
In the case of nested multiplications (like
\texttt{mul(mul(E1,E2),mul(E3,E4)))}), it would lead to naive re-evaluation
(of \texttt{E1}--\texttt{E4}), with a quadratic runtime in the worst case.
\end{minipage}
\end{tabular} \\[3pt]
Thanks to \texttt{symb/4}'s dual numbers, the expression is evaluated
incrementally, reusing intermediate results. This way the runtime
remains linear in the size of the expression.
\section{Gradients in Forward Mode}

Our previous definition of forward mode computes a single partial derivative
$\partialfrac{E}{X}$ at a time. However, often we are interested in computing
the \emph{gradient} $\nabla E$, which is the vector of partial derivatives with
respect to all variables.
To compute the gradient, we can repeatedly invoke \texttt{fwdad/4}, but this
is rather wasteful as it repeatedly computes the same primal for each partial
derivative. A more efficient approach computes the whole gradient in one go.
To accomplish that, we switch from the conventional dual numbers where
both the primal and tangent are numeric values to a more heterogeneous
structure where the tangent is the whole gradient\footnote{The generalized
dual number structure is known as Nagata's \emph{idealization of a module}
(\cite{DBLP:journals/corr/abs-2212-11088,Nagata1962}).}.

We represent the gradient by means of the common \texttt{assoc} library\footnote{
We have extended the library with \texttt{union\_with\_assoc/4} and
\texttt{insert\_with\_assoc/5}:
established AVL operations of respectively $\mathcal{O}(n)$ and $\mathcal{O}(n \log n)$
time complexity.
}
(based on AVL trees) as a partial map from variables $X$ to the partial
derivative $\partialfrac{E}{X}$. If a variable is not present in the map,
we assume its corresponding derivative is zero. This is convenient as
many intermediate results tend to be zero and thus need not be represented
explicitly.

\begin{Verbatim}[frame=single,fontsize=\small]
fwdadgrad(lit(N),_Env,N,DF) :-
  empty_assoc(DF).
fwdadgrad(var(Y),Env,F,DF) :-
  lookup(Y,Env,F),
  singleton_assoc(Y,1,DF).
fwdadgrad(add(E1,E2),Env,F,DF) :-
  fwdadgrad(E1,Env,F1,DF1),
  fwdadgrad(E2,Env,F2,DF2),
  F is F1 + F2,
  union_with_assoc(plus,DF1,DF2,DF).
fwdadgrad(mul(E1,E2),Env,F,SDF) :-
  fwdadgrad(E1,Env,F1,DF1),
  fwdadgrad(E2,Env,F2,DF2),
  F is F1 * F2,
  map_assoc(times(F2),DF1,SDF1),
  map_assoc(times(F1),DF2,SDF2),
  union_with_assoc(plus,SDF1,SDF2,SDF).

  plus(A,B,C) :- C is A + B.
  times(A,B,C) :- C is A * B.
\end{Verbatim}

If the expression tree has $N$ nodes and contains $V$ variables, then
\texttt{fwdad/4} has an $\mathcal{O}(NV)$ time complexity: each node is visited
once and the most costly operations \texttt{map\_assoc/3} and
\texttt{union\_with\_assoc/4} are both linear in $V$.

For a large number of variables $V$, the reverse-mode variant of AD
appears more efficient, but also more sophisticated.
In what follows we show how to obtain this reverse-mode AD by means of successively
optimizing scalar multiplication and vector addition.
This way, we become a declarative version of reverse-mode AD,
with an $\mathcal{O}(N\log{V})$ time complexity,
and its imperative counterpart, at $\mathcal{O}(N + V)$.

\section{Scalar Multipliers}

The first step towards efficient reverse-mode AD is to replace the costly
scalar multiplication of the gradient vector in the \texttt{mul/1} operation
with a constant-time multiplication.

Assume a scalar factor $M$ that should be multiplied with a gradient.
Instead of applying the scalar factor to the gradient vector in the
\texttt{mul/1} case, it is propagated down into the subexpressions. At the
\texttt{var/1} and \texttt{lit/0} leaves it becomes trivial to take the factor
$M$ into account: in the former case we create a single $M \times 1 = M$ vector entry;
in the latter case we ignore it as $M \times 0 = 0$ for all
variables and $0$ need not be explicitly represented.

In the case of nested multiplications, we pass multiple
factors $M_1$, \ldots $M_n$ down a path of the tree, which
can be combined into a single factor $M = M_1 \times \ldots \times M_n$.
In the absence of a factor we take $M = 1$ (the neutral element of
multiplication).

\begin{Verbatim}[frame=single,fontsize=\small]
revad1(E,Env,F,DF) :-                    revad1(mul(E1,E2),Env,M,F,DF) :-       
  revad1(E,Env,1,F,DF).                    times(M,F2,M1),
                                           times(M,F1,M2),
revad1(lit(N),_Env,_M,N,DF) :-             revad1(E1,Env,M1,F1,DF1),
  empty_assoc(DF).                         revad1(E2,Env,M2,F2,DF2),
revad1(var(Y),Env,M,F,DF) :-               F is F1 * F2,
  lookup(Y,Env,F),                         union_with_assoc(plus,DF1,DF2,DF).
  singleton_assoc(Y,M,DF).               
revad1(add(E1,E2),Env,M,F,DF) :-         
  revad1(E1,Env,M,F1,DF1),               
  revad1(E2,Env,M,F2,DF2),               
  F is F1 + F2,                          
  union_with_assoc(plus,DF1,DF2,DF).
\end{Verbatim}

There is one major snag in the above formulation. The modes in the
\texttt{mul/2} case do not work out: we need \texttt{F1} (to compute
\texttt{M2}), which is needed to compute \texttt{F2} and vice versa.
Fortunately, the mutual dependency is only an
artefact as the algorithm is set up as a single traversal of the
expression tree. In fact, \texttt{F1} does not depend on \texttt{F2}, only
\texttt{DF1} does. Hence, traditional algorithms work in two phases: they
compute the factors in a first phase and the derivatives in a second
phase. This requires additional book-keeping to store the appropriate
information at different nodes in the tree during the first phase for later
use in the second phase.

With Prolog's co-routine mechanisms, this problem is much easier to solve; it
does not require any restructuring of the code. We only need to revise the
definitions of \texttt{plus/3} and \texttt{times/3} to delay the arithmetic
operations until their inputs are known. We use \texttt{freeze(N,Goal)} for
this purpose, which defers the execution of \texttt{Goal} until \texttt{N} is
bound.

\begin{Verbatim}[frame=single,fontsize=\small]
plus(A,B,C) :- freeze(A,freeze(B,C is A + B)).
times(A,B,C) :- freeze(A,freeze(B,C is A * B)).
\end{Verbatim}

\section{Gradient Threading}

After eliminating the costly scalar multiplication of the gradient vector, we
now address the remaining costly operation: vector addition.

Vector additions combine the vectors that are created at the leaves of the
expression in a bottom-up fashion.
We change this dataflow to one that threads the
gradient vector through the computation and inserts elements where
they are created (in the \texttt{var/1} case). To accomplish the threading we make
use of Prolog's DCG notation (with association trees instead of lists),
which allows us to implicitly pass the gradient.
We initialize this vector with the neutral element of vector addition,
i.e., the empty vector.

\begin{Verbatim}[frame=single,fontsize=\small]
revad2(E,Env,F,DF) :-                    revad2(add(E1,E2),Env,M,F) --> 
  empty_assoc(DF0),                        revad2(E1,Env,M,F1),
  revad2(E,Env,1,F,DF0,DF).                revad2(E2,Env,M,F2),
                                           { F is F1 + F2 }.
revad2(lit(N),_Env,_M,F) -->             revad2(mul(E1,E2),Env,M,F) -->
  { F = N }.                               { times(M,F2,M1) },
revad2(var(Y),Env,M,F) -->                 { times(M,F1,M2) },
  { lookup(Y,Env,F) },                     revad2(E1,Env,M1,F1),
  insert_with_assoc(plus,Y,M).             revad2(E2,Env,M2,F2),
                                           { F is F1 * F2 }.
\end{Verbatim}

The resulting algorithm has $\mathcal{O}(N\log{V)}$ time complexity,
where the logarithmic factor is due to \texttt{insert\_with/3}.
We bring this down to $\mathcal{O}(1)$ by switching to destructive updates.

\section{Reverse-Mode AD, Destructively}\label{sec:destructive}

This section replaces the AVL tree representation of the gradient vector with
an array to achieve an overall $\mathcal{O}(N + V)$ time complexity.
The two operations we need to replace are the initialization of the gradient and
the insertion of a new value. Initialization goes from $\mathcal{O}(1)$ for
an empty AVL tree to $\mathcal{O}(V)$ to create a new array with an explicit 0
entry for each variable. Luckily, this operation is only performed once, at the start
of the algorithm. Insertion becomes a constant time operation thanks to the
\texttt{setarg/3} destructive update.
\begin{Verbatim}[frame=single,fontsize=\small]
empty_array(N,Arr) :-                   insert_with_array(Pred,Var,Value,Arr) :-
  findall(0,between(1,N,_),List),         arg(Var,Arr,OldValue),
  Arr =.. [grad|List].                    call(Pred,OldValue,Value,NewValue),
                                          setarg(Var,Arr,NewValue).
\end{Verbatim}

The impact of this representation change on the algorithm's
code itself is minimal.
\begin{Verbatim}[frame=single,fontsize=\small]
revad(E,Env,F,DF) :-                    revad(add(E1,E2),Env,M,F,DF) :-
  functor(Env,_,N),                       revad(E1,Env,M,F1,DF),
  empty_array(N,DF),                      revad(E2,Env,M,F2,DF),
  revad(E,Env,1,F,DF).                    F is F1 + F2.
                                        revad(mul(E1,E2),Env,M,F,DF) :-
revad(lit(N),_Env,_M,F,_DF) :-            times(M,F2,M1),
  F = N.                                  times(M,F1,M2),
revad(var(Y),Env,M,F,DF) :-               revad(E1,Env,M1,F1,DF),
  lookup(Y,Env,F),                        revad(E2,Env,M2,F2,DF),
  insert_with_array(plus,Y,M,DF).         F is F1 * F2.
\end{Verbatim}

\section{Extensions}\label{sec:extensions}

So far, we have only considered a minimal grammar for expressions. Many
applications of AD involve additional mathematical functions (e.g., trigonometric,
exponential, logarithmic), which can be easily incorporated into
our approach.
For example, we can extend our expression language with negation \texttt{neg(E)},
sine \texttt{sin(E)} and cosine \texttt{cos(E)}, exponentials \texttt{exp(E)}, \ldots.
The chain rule explains how to support extensions:
\begin{equation*}
\partialfrac{f(e)}{x} = \partialfrac{f(e)}{e} \times \partialfrac{e}{x}
\end{equation*}
where $\partialfrac{f(e)}{e}$ is the derivative of $f$ with respect to its argument
and evaluated at $e$.
The following clauses incorporate these additional functions
in the last version we have presented (left) based on the symbolic derivatives
of the functions (right).

\noindent
\begin{tabular}{ll}
\begin{minipage}{.6\textwidth}
\begin{Verbatim}[frame=single,fontsize=\small]
revad(neg(E1),Env,M,F,DF) :-
  freeze(M,freeze(F1,M1 is M * -F1)),
  revad(E1,Env,M1,F1,DF),
  F is -F1.
revad(sin(E1),Env,M,F,DF) :-
  freeze(M,freeze(F1,M1 is M * cos(F1))),
  revad(E1,Env,M1,F1,DF),
  F is sin(F1).
revad(cos(E1),Env,M,F,DF) :-
  freeze(M,freeze(F1,M1 is M * -sin(F1))),
  revad(E1,Env,M1,F1,DF),
  F is cos(F1).
revad(exp(E1),Env,M,F,DF) :-
  freeze(M,freeze(F1,M1 is M * exp(F1))),
  revad(E1,Env,M1,F1,DF),
  F is exp(F1).
\end{Verbatim}
\end{minipage}
&
\begin{minipage}{.4\textwidth}
\small
\begin{eqnarray*}
\partialfrac{-x}{x} & = & -x \\  \\
\partialfrac{\sin{x}}{x} & = & \cos{x} \\ \\
\partialfrac{\cos{x}}{x} & = & -\sin{x} \\  \\
\partialfrac{\exp{x}}{x} & = & \exp{x}
\end{eqnarray*}
\end{minipage}
\end{tabular}

The chain rule generalizes to multi-argument functions as follows:
\begin{equation*}
\partialfrac{f(e_1,\ldots,e_n)}{x} = \sum_{i=1}^n \partialfrac{f(e_1,\ldots,e_n)}{e_i} \times \partialfrac{e_i}{x}
\end{equation*}
where $f_i'$ is the partial derivative of $f$ with respect to its $i$th argument.

For example, we use the notation \texttt{pow(E$_1$,E$_2$)} to denote \texttt{E$_1$} raised
to the power \texttt{E$_2$}.

\noindent
\begin{tabular}{ll}
\begin{minipage}{.6\textwidth}
\begin{Verbatim}[frame=single,fontsize=\small]
revad(pow(E1,E2),Env,M,F,DF) :-
  freeze(M,freeze(F1,freeze(F2,
      M1 is M * F2 * F1**(F2 - 1)))),
  freeze(M,freeze(F1,freeze(F2,
      M2 is M * (F1**F2) * log(F1)))),
  revad(E1,Env,M1,F1,DF),
  revad(E2,Env,M2,F2,DF),
  F is F1**F2.
\end{Verbatim}
\end{minipage}
&
\begin{minipage}{.4\textwidth}
\small
\begin{eqnarray*}
\partialfrac{x^b}{x} & = & bx^{b-1} \\  \\
\partialfrac{a^x}{x} & = & a^x\ln{a}
\end{eqnarray*}
\end{minipage}
\end{tabular}
\section{Case Study: Sum-Product Loop Programming}\label{sec:spll}

\citet{DBLP:conf/kr/PfanschillingSD22} have recently proposed the Sum-Product
Loop Language (SPLL), but (as far as we are aware) no implementation has been
made publicly available. As a case study, we have implemented this probabilistic
language in Prolog, and notably its parameter estimation functionality,
with our AD algorithms.

\paragraph{Parameter Estimation}

An SPLL program defines a probability distribution over its possible outcomes.
For instance, \Verb!main = Uniform >= Theta[1]!
transforms the \Verb!Uniform! distribution on the interval $[0,1]$ into a Bernoulli distribution:
$p(\texttt{false}) = \theta_1$ and $p(\texttt{true}) = 1 - \theta_1$.
Formally, SPLL comes with both a generative and a probabilistic
semantics. The former samples the program (i.e., generates a result $x$) in
accordance with its probability distribution, and the latter computes the
probability $p(x)$ of a given sample $x$. The probabilistic semantics can be used to estimate the
parameters $\theta$ of a program in terms of a set of samples $X$. Specifically,
given a set of samples $X$ and a consistent SPLL program, find the parameterization
$\theta$ of the program that minimizes the negative log-likelihood $\mathcal{L}$ of all samples:
$\mathcal{L} = \sum_{x \in X} -\log{p(x|\theta)}$.

This problem can be tackled with gradient-based optimization, where the
gradient $\partialfrac{\mathcal{L}}{\theta}$ is computed with automatic
differentiation. Indeed, given a learning rate $\lambda$ and an initial guess
for the parameters $\theta$, we can iteratively improve the parameters with
$\theta := \theta - \lambda \partialfrac{L}{\theta}$.
We have implemented a symbolic version of the probabilistic semantics, which
yields
for our small example program

\noindent
\begin{minipage}{0.5\textwidth}
  \begin{eqnarray*}
  p(\texttt{true})  & = & \int_{\theta_1}^{\infty} \varphi_U(x)\,dx
  \end{eqnarray*}
\end{minipage}%
\begin{minipage}{0.5\textwidth}
  \begin{eqnarray*}
  p(\texttt{false}) & = & 1 - \int_{\theta_1}^{\infty} \varphi_U(x)\,dx
  \end{eqnarray*}
\end{minipage}

Given an initial guess $\theta_1 = 0.5$, a learning rate of $0.02$, and
a set $X$ of
3 \texttt{false} samples and 7 \texttt{true} samples, the parameter
converges to 0.3000000000000001 in 13 steps.

\paragraph{Benchmarking our AD Versions}

To compare the runtime of the different AD versions we have 
used the following SPLL program, which features six parameters $\theta_1$,\ldots,$\theta_6$.

\begin{verbatim}
main = if Uniform >= Theta[1]
       then if Uniform >= Theta[2]
            then if Uniform >= Theta[3] then null    else [true]
            else if Uniform >= Theta[4] then [false] else [true,true]
       else if Uniform >= Theta[5]
            then if Uniform >= Theta[6] then [true,false] else [false,true]
            else [false,false]
\end{verbatim}

We have used the same gradient-based optimization of the negative log-likelihood to simultaneously learn
the six parameters.
Given an initial guess of $0.5$ for $\theta_1$ and $0.25$ for the others, a learning rate
of $0.02$ and sets of 3 examples for each outcome, the parameters converge to the following
in 100 iterations:

  \begin{center}
  \ra{1.3}
    \begin{tabular}{@{}lll@{}}\toprule
      \textbf{Parameter} & \textbf{Learned Value} & \textbf{Optimal Value} \\ \midrule
      $\theta_1$ & $0.4285714285714287$  & $3/7$ \\
      $\theta_2$ & $0.5$                 & $1/2$ \\
      $\theta_3$ & $0.49999999999999994$ & $1/2$ \\
      $\theta_4$ & $0.49999999999999994$ & $1/2$ \\
      $\theta_5$ & $0.3333333333333333$  & $1/3$ \\
      $\theta_6$ & $0.49999999999999994$ & $1/2$ \\ 
      \bottomrule
    \end{tabular}
  \end{center}
We have also measured the runtime of this experiment (repeated 100 times) using
our four different AD algorithms.
These measurements were performed on a Macbook Pro with M1 chip, 16GB RAM, Ventura 13.1,
with SWI-Prolog version 8.4.1 for arm64-darwin.
  \begin{center}
  \ra{1.3}
    \begin{tabular}{@{}rrrrr@{}}\toprule
      \textbf{AD version}  & \texttt{fwdadgrad} & \texttt{revad1} &  \texttt{revad1} & \texttt{revad}  \\ \midrule
      \textbf{Runtime (s)} & $9.346$            & $9.098$          & $5.768$          & $4.909$  \\
    \bottomrule
    \end{tabular}
  \end{center}
These results demonstrate the improvements achieved by the successive algorithms.

\section{Case Study: Probabilistic Logic Programming}\label{sec:plp}
\label{plp}
Automatic differentiation has various applications in Probabilistic Logic
Programming \citep{Rig18-BKaddress}.  For instance, it can be used to
perform both parameter learning and variational inference on \emph{hybrid}
programs, which are programs that include both continuous and discrete random
variables.

\subsection{Learning Hybrid Programs}
\cite{TLP:8688161} proposed Extended PRISM, a version of PRISM \citep{sato1997prism} that allows continuous random variables with a
Gaussian or gamma distribution.

PRISM introduces random atoms via the predicate
\verb|msw/2| where the first argument is a {\em random switch name}, a
term representing a discrete random variable, and the second argument
represents a value for that variable.
An example atom is
\verb|msw(m,a)|.

The set of possible values for
a switch is defined by a fact for the predicate \verb|values/2| where the first argument is the
name of the switch and the second argument is a list of terms representing its possible values.
For example, \texttt{values(m,[a,b])}.

The probability distribution over the values of the random variable
associated with a switch name is defined by a directive for the predicate \verb|set_sw/2| where the first argument is the name
of the switch and the second argument is a list of probabilities. For example:
\begin{verbatim}
:- set_sw(m, [0.3, 0.7]).
\end{verbatim}
The semantics of the language can be defined by specifying a way to sample values for the variables of a query atom
from the program: the query atom is answered by resolution for deterministic programs with the only difference that, each time a
\verb|msw(name,value)| atom is encountered, a value is sampled from the distribution for \verb|name| and unified with \verb|value|.
Extended PRISM adds to PRISM continuous random variables with a Gaussian or gamma distribution.
In this case the \verb|values/2| predicate has \verb|real| instead of the list of values and
the directive \verb|set_sw/2| specifies the probability density such as
\verb|norm(0.5,0.1)| for a Gaussian distribution with mean 0.5 and variance 0.1.

Let us illustrate the language with an example.
Suppose a factory has two machines $a$ and $b$. Each machine produces a widget
with a continuous feature. A widget is produced by machine $a$ with probability
0.3 and by machine $b$ with probability $0.7$.
If the widget is produced by machine $a$, the continuous feature is distributed as a
Gaussian with mean 2.0 and variance 1.0.
If the widget is produced by machine $b$, the continuous feature is distributed as a
Gaussian with mean 3.0 and variance 1.0.
A third machine then adds a random quantity to
the feature. The quantity is distributed as a Gaussian with mean 0.5 and variance 0.1.
This is encoded by the program:

\begin{Verbatim}[frame=single,fontsize=\small]
widget(X) :- msw(m, M), msw(st(M), Z), msw(pt, Y), X = Y + Z.
values(m, [a,b]).
values(st(_), real).
values(pt, real).
:- set_sw(m, [0.3, 0.7]).
:- set_sw(st(a), norm(2.0, 1.0)).
:- set_sw(st(b), norm(3.0, 1.0)).
:- set_sw(pt, norm(0.5, 0.1)).
\end{Verbatim}
\cite{TLP:8688161} presented an inference algorithm that solves the DISTR task \citep{Rig18-BKaddress}:
computing the probability distribution or density of the non-ground arguments of a conjunction of literals, e.g., computing the probability density of \verb|X| in goal \verb|widget(X)| of the example above.
The algorithm collects symbolic derivations for the query and then builds a representation of the probability density associated with the variables of each goal, bottom-up, starting from the leaves.

For the widget example, the probability density of \verb|X| in goal \verb|widget(X)| is \citep{TLP:8688161}:
$$p(x)=0.3\cdot \varphi_N(x;2.5,1.1)+0.7\cdot \varphi_N(x;3.5,1.1)$$
with $\varphi_N(x;\mu,\sigma^2)$ the density of a Gaussian distribution with mean $\mu$ and variance $\sigma^2$:
$$\varphi_N(x; \mu,\sigma^2) = \frac{1}{\sigma \sqrt{2\pi} } e^{-\frac{1}{2}\left(\frac{x-\mu}{\sigma}\right)^2}$$
For Extended PRISM, the problem of parameter learning consists of being given a set of ground atoms $X$ for a query predicate and a program with some unknown parameters
and finding the parameters that maximize the log-likelihood of the atoms in $X$.

This problem can be solved by using inference to find the density of the arguments of the query predicate, using automatic differentiation for finding the derivatives of the
density with respect to the various parameters and then using gradient ascent for optimizing the log-likelihood, similarly to SPLL.
For the widget example,  the parameters  could be learned
with this approach from a set of ground atoms for the predicate \verb|widget/1|.

We performed an experiment that learns back the parameters for the \verb|pt| switch from data.
To do so we generated 50,000 samples of the query \verb|widget(X)| from the program above, resulting in 50,000 values for \verb|X|.
Then we replaced
\begin{verbatim}
:- set_sw(pt, norm(0.5, 0.1)).
\end{verbatim}
with
\begin{verbatim}
:- set_sw(pt, norm(mu, sigma2)).
\end{verbatim}
and we generated with inference the probability density for \verb|X| in \verb|widget(X)|, obtaining 
$$p(x)=0.3\cdot \varphi_N(x;2+\mu,1+\sigma^2)+0.7\cdot \varphi_N(x;3+\mu,1+\sigma^2)$$
where $\mu$ stands for \verb|mu| and $\sigma^2$ for \verb|sigma2|.
We applied gradient-based optimization as for the SPLL case. To make sure $\sigma^2$ remains positive
we reparametrized it using a weight $W$ between $-\infty$ and $+\infty$ and expressing $\sigma^2$ as $e^W$.
We replace $\sigma^2$ in the formula with $e^W$ and we computed the derivatives of $p(x)$ with respect to $\mu$ and $W$ using
\verb|revad/5|. Once the procedure terminates and returns the values for $\mu$ and $W$, we obtain $\sigma^2$ as $e^W$.

With a  learning rate of 0.00005, the procedure converged in 29 iterations giving  the values 0.502130 for $\mu$ and 0.002331 for $\sigma^2$ that are quite close to the true values.
The procedure took 153.27 seconds on a 2.8 GHz Intel Core i7 using SWI-Prolog.
\subsection{Variational Inference}
In variational inference we want to approximate a difficult-to-compute conditional probability density $p(x|y)$ with a simpler distribution $p_\theta(x)$.
In the widget example, we may want to compute $p(Y|X=0.2)$ where $X$ and $Y$ are the variables appearing in the program as \verb|X| and \verb|Y|.
The simpler distribution  $p_\theta(x)$ is parameterized by a set of parameters $\theta$ and we want to optimize them in order to make
  $p_\theta(x)$ as similar as possible to $p(x|y)$.
This is typically done by minimizing
the  Kullback–Leibler divergence:
\begin{eqnarray}
\label{eq:kl}
KL(p_\theta(x),p(x|y)) & = & \int_x p_{\theta}( x ) \log \left(
  \frac{p_\theta(x)}{p(x|y)} \right) \nonumber \\
& = & \int_x p_{\theta}( x ) \log \left(
  \frac{p_\theta(x)}{p(y|x)p(x)} \right) + \log p(y) \nonumber \\
& = & -L(\theta)+\log p(y)\\
  \text{where} \nonumber \\
   L(\theta)& \overset{\Delta}{=}& \int_x p_{\theta}( x ) \log \left(
  \frac{p(y|x)p(x)}{p_\theta(x)} \right)
\end{eqnarray}
Since $\log p(y)$ does not depend on $\theta$, the divergence is minimized by maximizing $L(\theta)$.
The maximization can be performed using gradient ascent if we can compute the derivatives of $L(\theta)$ in terms of $\theta$.
\cite{https://doi.org/10.48550/arxiv.1301.1299} proposed an approach to apply variational inference to probabilistic programming in general.
The gradient can be estimated according to the following formula \citep{https://doi.org/10.48550/arxiv.1301.1299}
\begin{equation}
-\nabla_\theta\: L(\theta)= \approx \frac{1}{N}  \sum_{x^j} \nabla_{\theta} \log p_{\theta}( x^j
)\left(  \log \left( \frac{ p_{\theta}( x^j ) }{ p(y|x^j)p(x^j)} \right)+K \right)\label{eq4}
\end{equation}
which is  a Monte Carlo approximation of an integral where $x^j \sim p_{\theta}( x )$, $j=1\ldots N$ and $K$ is an arbitrary constant.

Suppose $p(x)$ is given by a probabilistic (logic) program. Our aim is to find a \emph{target program} encoding the conditional distribution $p(x|y)$. We do so
by considering another program encoding $p_\theta(x)$ that we call the \emph{variational program}: we optimize the parameters $\theta$ to make $p_\theta(x)$ as
similar as possible to $p(x|y)$. We use inference to compute $p_\theta(x)$ for the variational program and we use automatic differentiation to compute $ \nabla_{\theta} \log p_{\theta}( x)$.
From equation (\ref{eq4}) we see that what is left to do is to compute $p(y|x^j)$ which is easy to execute with forward inference: in the widget example it means computing $p(X|Y)$ where 
forward inference can be applied after setting $Y$ to a fixed value.
\section{Related Work}\label{sec:related}

There has been a lot of work on automatic differentiation in the general
programming languages research community in recent years~(e.g., \cite{pacmpl/KrawiecJKEEF22,pacmpl/WangZDWER19,higher-order,popl2023,DBLP:journals/toplas/PearlmutterS08}). Much of this work has been
focused on extending the expressivity of AD (e.g., to higher-order functions)
and on ways of showing the correctness of different AD flavors. Several
works also focus on the compositionality of models (\cite{DBLP:journals/pacmpl/NguyenPWW22}, \cite{DBLP:journals/pacmpl/DashKPS23}).

This paper is
most closely related to that of \cite{DBLP:journals/corr/abs-2212-11088}, which
provides a (Haskell-based) account of AD in terms of various forms of
generalized dual numbers. A key difference with these existing works is that they
use ``Wengert lists'', ``tapes'' or function abstractions to defer computations
of tangents that depend on primals, while our Prolog approach can simply use coroutines.

In contrast, as far as we know, there is little work on AD in the logic programming
community. We have found only two (unpublished) manuscripts.
The first is a paper by~\cite{Samer} on implementing AD with Constraint
Handling Rules (CHR), which involves a different programming style (e.g., using
constraints to represent the expression AST and using delay mechanisms much more
extensively) and deviates more from the traditional algorithms. Moreover, it
presents two algorithms as given,
rather than deriving them systematically like we do.
The second is an unfinished blogpost by~\cite{Gabel}, which provides a Prolog
implementation of forward-mode AD that takes a compilation-based approach
and produces a sequence of assignments.

Algebraic ProbLog (\cite{aproblog}) uses semirings to support automatic differentiation in the domain of 
probabilistic logic programming,
which is used by DeepProbLog (\cite{deepproblog}) for gradient-descent optimization in their backpropagation.

Many machine learning libraries come with highly optimized automatic
differentiation implementations (e.g., with tailored memory management
techniques and GPU leverage), such as Tensorflow's (\cite{tensorflow2015}) and PyTorch's
(\cite{pytorch}) Autograd, which targets Python code. Google's JAX (\cite{jax}) extends
Autograd with just-in-time compilation, inspired by Tensorflow's XLA
(accelerated linear algebra). The probabilistic programming language Stan
(\cite{stan,carpenter2015stan}) leverages C++ to implement automatic
differentiation. 
All of these easily outperform our Prolog implementations, whose aim is not to
provide competitive performance. Instead, we aim to provide an account of AD
for the logic programming community that is instructive, accessible and
lowers the threshold for experimentation.

\section{Conclusion}\label{sec:conclusion}

We have shown that forward-mode and reverse-mode automatic differentiation in
Prolog can be systematically derived from their specification in terms of
symbolic differentiation and evaluation. Their definitions are elegant and
concise and achieve the textbook theoretical time complexities.

In future work we plan to explore Prolog's meta-programming facilities (e.g.,
term expansion) to implement partial evaluation of \texttt{revad/5} calls on
known expressions. We also wish to develop further applications on top of our
AD approach, such as Prolog-based neural networks and integration with
existing probabilistic logic programming languages.

\bibliographystyle{tlplike}
\bibliography{bib}

\begin{thebibliography}{}

\bibitem[Abadi et~al., 2015]{tensorflow2015}
{\sc Abadi, M.} {\sc and} {\sc others} 2015.
\newblock {TensorFlow}: Large-scale machine learning on heterogeneous systems.
\newblock Software available from tensorflow.org.

\bibitem[Abdallah, 2017]{Samer}
{\sc Abdallah, S.} 2017.
\newblock Automatic differentiation using constraint handling rules in
  {P}rolog.
\newblock \url{https://arxiv.org/abs/1706.00231}.

\bibitem[Bradbury et~al., 2018]{jax}
{\sc Bradbury, J.}, {\sc Frostig, R.}, {\sc Hawkins, P.}, {\sc Johnson, M.~J.},
  {\sc Leary, C.}, {\sc Maclaurin, D.}, {\sc Necula, G.}, {\sc Paszke, A.},
  {\sc Vander{P}las, J.}, {\sc Wanderman-{M}ilne, S.}, {\sc and} {\sc Zhang,
  Q.} 2018.
\newblock {JAX}: composable transformations of {P}ython+{N}um{P}y programs.

\bibitem[Bruynooghe, 2004]{DBLP:journals/tplp/Bruynooghe04}
{\sc Bruynooghe, M.} 2004.
\newblock Enhancing a search algorithm to perform intelligent backtracking.
\newblock {\em Theory Pract. Log. Program.}, {\it 4}, 3, 371--380.

\bibitem[Carpenter et~al., 2017]{stan}
{\sc Carpenter, B.}, {\sc Gelman, A.}, {\sc Hoffman, M.~D.}, {\sc Lee, D.},
  {\sc Goodrich, B.}, {\sc Betancourt, M.}, {\sc Brubaker, M.}, {\sc Guo, J.},
  {\sc Li, P.}, {\sc and} {\sc Riddell, A.} 2017.
\newblock Stan: A probabilistic programming language.
\newblock {\em Journal of statistical software}, {\it 76}, 1.

\bibitem[Carpenter et~al., 2015]{carpenter2015stan}
{\sc Carpenter, B.}, {\sc Hoffman, M.~D.}, {\sc Brubaker, M.}, {\sc Lee, D.},
  {\sc Li, P.}, {\sc and} {\sc Betancourt, M.} 2015.
\newblock The stan math library: Reverse-mode automatic differentiation in c++.
\newblock {\em arXiv preprint arXiv:1509.07164},.

\bibitem[Clocksin and Mellish, 2003]{ClocksinMellish03}
{\sc Clocksin, W.~F.} {\sc and} {\sc Mellish, C.~S.} 2003.
\newblock {\em Programming in Prolog}.
\newblock Springer, Berlin, 5 edition.

\bibitem[Dash et~al., 2023]{DBLP:journals/pacmpl/DashKPS23}
{\sc Dash, S.}, {\sc Kaddar, Y.}, {\sc Paquet, H.}, {\sc and} {\sc Staton, S.}
  2023.
\newblock Affine monads and lazy structures for bayesian programming.
\newblock {\em Proc. {ACM} Program. Lang.}, {\it 7}, {POPL}, 1338--1368.

\bibitem[Debray, 1988]{10.1145/53990.54020}
{\sc Debray, S.~K.}
\newblock Unfold/fold transformations and loop optimization of logic programs.
\newblock In {\em Proceedings of the ACM SIGPLAN 1988 Conference on Programming
  Language Design and Implementation} 1988, PLDI '88,  297–307, New York, NY,
  USA.

\bibitem[Gabel, 2020]{Gabel}
{\sc Gabel, P.~L.} 2020.
\newblock Automatic differentiation.
\newblock
  \url{http://petergabel.info/blog/AutoDiff/#Backward-Mode-Automatic-Differentiation-1}.

\bibitem[Howe and King, 2012]{DBLP:journals/tcs/HoweK12}
{\sc Howe, J.~M.} {\sc and} {\sc King, A.} 2012.
\newblock A pearl on {SAT} and {SMT} solving in prolog.
\newblock {\em Theor. Comput. Sci.}, {\it 435}, 43--55.

\bibitem[Islam et~al., 2012]{TLP:8688161}
{\sc Islam, M.~A.}, {\sc Ramakrishnan, C.}, {\sc and} {\sc Ramakrishnan, I.}
  2012.
\newblock Inference in probabilistic logic programs with continuous random
  variables.
\newblock {\em Theory Pract. Log. Program.}, {\it 12}, 505--523.

\bibitem[Kimmig et~al., 2011]{aproblog}
{\sc Kimmig, A.}, {\sc Van~den Broeck, G.}, {\sc and} {\sc De~Raedt, L.}
\newblock An algebraic prolog for reasoning about possible worlds.
\newblock In {\em Twenty-Fifth AAAI Conference on Artificial Intelligence}
  2011.

\bibitem[Kowalski, 1979]{DBLP:journals/cacm/Kowalski79}
{\sc Kowalski, R.~A.} 1979.
\newblock Algorithm = logic + control.
\newblock {\em Commun. {ACM}}, {\it 22}, 7, 424--436.

\bibitem[Krawiec et~al., 2022]{pacmpl/KrawiecJKEEF22}
{\sc Krawiec, F.}, {\sc Jones, S.~P.}, {\sc Krishnaswami, N.}, {\sc Ellis, T.},
  {\sc Eisenberg, R.~A.}, {\sc and} {\sc Fitzgibbon, A.~W.} 2022.
\newblock Provably correct, asymptotically efficient, higher-order reverse-mode
  automatic differentiation.
\newblock {\em Proc. {ACM} Program. Lang.}, {\it 6}, {POPL}, 1--30.

\bibitem[Manhaeve et~al., 2018]{deepproblog}
{\sc Manhaeve, R.}, {\sc Dumancic, S.}, {\sc Kimmig, A.}, {\sc Demeester, T.},
  {\sc and} {\sc De~Raedt, L.} 2018.
\newblock Deepproblog: Neural probabilistic logic programming.
\newblock {\em advances in neural information processing systems}, {\it 31}.

\bibitem[Nagata, 1962]{Nagata1962}
{\sc Nagata, M.} 1962.
\newblock {\em Local rings}.
\newblock Wiley Interscience.

\bibitem[Nguyen et~al., 2022]{DBLP:journals/pacmpl/NguyenPWW22}
{\sc Nguyen, M.}, {\sc Perera, R.}, {\sc Wang, M.}, {\sc and} {\sc Wu, N.}
  2022.
\newblock Modular probabilistic models via algebraic effects.
\newblock {\em Proc. {ACM} Program. Lang.}, {\it 6}, {ICFP}, 381--410.

\bibitem[Paszke et~al., 2019]{pytorch}
{\sc Paszke, A.}, {\sc Gross, S.}, {\sc Massa, F.}, {\sc Lerer, A.}, {\sc
  Bradbury, J.}, {\sc Chanan, G.}, {\sc Killeen, T.}, {\sc Lin, Z.}, {\sc
  Gimelshein, N.}, {\sc Antiga, L.}, {\sc Desmaison, A.}, {\sc Kopf, A.}, {\sc
  Yang, E.}, {\sc DeVito, Z.}, {\sc Raison, M.}, {\sc Tejani, A.}, {\sc
  Chilamkurthy, S.}, {\sc Steiner, B.}, {\sc Fang, L.}, {\sc Bai, J.}, {\sc
  and} {\sc Chintala, S.}
\newblock Pytorch: An imperative style, high-performance deep learning library.
\newblock In {\em Advances in Neural Information Processing Systems 32} 2019,
  pp. 8024--8035. Curran Associates, Inc.

\bibitem[Pearlmutter and Siskind, 2008]{DBLP:journals/toplas/PearlmutterS08}
{\sc Pearlmutter, B.~A.} {\sc and} {\sc Siskind, J.~M.} 2008.
\newblock Reverse-mode {AD} in a functional framework: Lambda the ultimate
  backpropagator.
\newblock {\em {ACM} Trans. Program. Lang. Syst.}, {\it 30}, 2, 7:1--7:36.

\bibitem[Pfanschilling et~al., 2022]{DBLP:conf/kr/PfanschillingSD22}
{\sc Pfanschilling, V.}, {\sc Shindo, H.}, {\sc Dhami, D.~S.}, {\sc and} {\sc
  Kersting, K.}
\newblock Sum-product loop programming: From probabilistic circuits to loop
  programming.
\newblock In {\em Proceedings of {KR}} 2022.

\bibitem[Riguzzi, 2018]{Rig18-BKaddress}
{\sc Riguzzi, F.} 2018.
\newblock {\em Foundations of Probabilistic Logic Programming: Languages,
  semantics, inference and learning}.
\newblock River Publishers, Gistrup, Denmark.

\bibitem[Robbins et~al., 2021]{DBLP:journals/tplp/RobbinsKH21}
{\sc Robbins, E.}, {\sc King, A.}, {\sc and} {\sc Howe, J.~M.} 2021.
\newblock Backjumping is exception handling.
\newblock {\em Theory Pract. Log. Program.}, {\it 21}, 2, 125--144.

\bibitem[Sato and Kameya, 1997]{sato1997prism}
{\sc Sato, T.} {\sc and} {\sc Kameya, Y.}
\newblock {PRISM}: a language for symbolic-statistical modeling.
\newblock In {\em IJCAI} 1997, volume~97, pp. 1330--1339.

\bibitem[Schrijvers and Fr{\"{u}}hwirth,
  2006]{DBLP:journals/tplp/SchrijversF06}
{\sc Schrijvers, T.} {\sc and} {\sc Fr{\"{u}}hwirth, T.~W.} 2006.
\newblock Optimal union-find in constraint handling rules.
\newblock {\em Theory Pract. Log. Program.}, {\it 6}, 1-2, 213--224.

\bibitem[Smeding and V\'{a}k\'{a}r, 2022]{popl2023}
{\sc Smeding, T.} {\sc and} {\sc V\'{a}k\'{a}r, M.} 2022.
\newblock Efficient dual-numbers reverse ad via well-known program
  transformations.
\newblock \url{https://arxiv.org/abs/2207.03418}.

\bibitem[Szirmay-Kalos, 2021]{higher-order}
{\sc Szirmay-Kalos, L.} 2021.
\newblock Higher order automatic differentiation with dual numbers.
\newblock {\em Periodica Polytechnica Electrical Engineering and Computer
  Science}, {\it 65}, 1, 1--10.

\bibitem[van~den Berg et~al., 2022]{DBLP:journals/corr/abs-2212-11088}
{\sc van~den Berg, B.}, {\sc Schrijvers, T.}, {\sc McKinna, J.}, {\sc and} {\sc
  Vandenbroucke, A.} 2022.
\newblock Forward- or reverse-mode automatic differentiation: What's the
  difference?
\newblock {\em CoRR}, {\it abs/2212.11088}.

\bibitem[Vandecasteele and Janssens, 2003]{DBLP:journals/tplp/VandecasteeleJ03}
{\sc Vandecasteele, H.} {\sc and} {\sc Janssens, G.} 2003.
\newblock An open ended tree.
\newblock {\em Theory Pract. Log. Program.}, {\it 3}, 3, 377--385.

\bibitem[Wang et~al., 2019]{pacmpl/WangZDWER19}
{\sc Wang, F.}, {\sc Zheng, D.}, {\sc Decker, J.~M.}, {\sc Wu, X.}, {\sc
  Essertel, G.~M.}, {\sc and} {\sc Rompf, T.} 2019.
\newblock Demystifying differentiable programming: shift/reset the penultimate
  backpropagator.
\newblock {\em Proc. {ACM} Program. Lang.}, {\it 3}, {ICFP}, 96:1--96:31.

\bibitem[Wengert, 1964]{Wengert64}
{\sc Wengert, R.~E.} 1964.
\newblock A simple automatic derivative evaluation program.
\newblock {\em Commun. ACM}, {\it 7}, 8, 463–464.

\bibitem[Wingate and Weber, 2013]{https://doi.org/10.48550/arxiv.1301.1299}
{\sc Wingate, D.} {\sc and} {\sc Weber, T.} 2013.
\newblock Automated variational inference in probabilistic programming.

\end{thebibliography}

\newpage
\appendix

\section{Derivation of \texttt{fwdad/4} from \texttt{ad\_spec/4}}
\label{app:fwdad}

\begin{Verbatim}
ad_spec(E,X,Env,N,DN) :-
  symb(E,X,F,DF),
  eval(F,Env,N),
  eval(DF,Env,DN).

<=> (unfold symb)

  * Case lit/1

      ad_spec(lit(M),X,Env,N,DN) :-
        lit(M),X,lit(N),lit(0)
        eval(lit(M),Env,N),
        eval(lit(0),Env,DN).

    <=> (unfold of eval)

      ad_spec(lit(M),X,Env,M,0).

  * Case var/1

      ad_spec(var(Y),X,Env,N,DN) :-
        ( X == Y -> DF = 1 ; DF = 0),
        eval(var(Y),Env,N),
        eval(lit(DF),Env,DN).

    <=> (commutativity of ,/2 and distributivity of ,/2 wrt (->;))

      ad_spec(var(Y),X,Env,N,DN) :-
        eval(var(Y),Env,N),
        ( X == Y -> eval(lit(1),Env,DN); eval(lit(0),Env,DN)).

    <=> (unfold eval)

      ad_spec(var(Y),X,Env,N,DN) :-
        lookup(Y,Env,N),
        ( X == Y -> DN = 1; DN = 0).

  * Case add/2

      ad_spec(add(E1,E2),X,Env,N,DN) :-
        symb(E1,X,F1,DF1),
        symb(E2,X,F2,DF2),
        eval(add(F1,F2),Env,N),
        eval(add(DF1,DF2),Env,DN).

    <=> (unfold eval)

      ad_spec(add(E1,E2),X,Env,N,DN) :-
        symb(E1,X,F1,DF1),
        symb(E2,X,F2,DF2),
        eval(F1,Env,N1),
        eval(F2,Env,N2),
        N is N1 + N2,
        eval(DF1,Env,DN1),
        eval(DF2,Env,DN2),
        DN is DN1 + DN2.

    <=> (commutativity of ,/2)

      ad_spec(add(E1,E2),X,Env,N,DN) :-
        symb(E1,X,F1,DF1),
        eval(F1,Env,N1),
        eval(DF1,Env,DN1),
        symb(E2,X,F2,DF2),
        eval(DF2,Env,DN2),
        eval(F2,Env,N2),
        N is N1 + N2,
        DN is DN1 + DN2.

    <=> (fold of ad_spec)

      ad_spec(add(E1,E2),X,Env,N,DN) :-
        ad_spec(E1,X,N1,DN1),
        symb(E2,X,N2,DN2),
        N is N1 + N2,
        DN is DN1 + DN2.

  * Case mul/2

      ad_spec(mul(E1,E2),X,Env,N,DN) :-
        symb(E1,X,F1,DF1),
        symb(E2,X,F2,DF2),
        eval(mul(F1,F2),Env,N),
        eval(add(mul(F1,DF2),mul(F2,DF1)),Env,DN).

    <=> (unfold eval)

      ad_spec(mul(E1,E2),X,Env,N,DN) :-
        symb(E1,X,F1,DF1),
        symb(E2,X,F2,DF2),
        eval(F1,Env,N1),
        eval(F2,Env,N2),
        N is N1 * N2,
        eval(F1,Env,M1),
        eval(DF2,Env,DN2),
        eval(F2,Env,M2),
        eval(DF1,Env,DN1),
        DN is M1 * DN2 + M2 * DN1.

    <=> (functional dependency of eval and idempotence)

      ad_spec(mul(E1,E2),X,Env,N,DN) :-
        symb(E1,X,F1,DF1),
        symb(E2,X,F2,DF2),
        eval(F1,Env,N1),
        eval(F2,Env,N2),
        N is N1 * N2,
        eval(DF2,Env,DN2),
        eval(DF1,Env,DN1),
        DN is N1 * DN2 + N2 * DN1.

    <=> (commutativity of ,/2)

      ad_spec(mul(E1,E2),X,Env,N,DN) :-
        symb(E1,X,F1,DF1),
        eval(F1,Env,N1),
        eval(DF1,Env,DN1),
        symb(E2,X,F2,DF2).
        eval(F2,Env,N2),
        eval(DF2,Env,DN2),
        N is N1 * N2,
        DN is N1 * DN2 + N2 * DN1.

    <=> (fold of ad_spec)

      ad_spec(mul(E1,E2),X,Env,N,DN) :-
        ad_spec(E1,X,N1,DN1),
        ad_spec(E2,X,N2,DN2),
        N is N1 * N2,
        DN is N1 * DN2 + N2 * DN1.
\end{Verbatim}

\end{document}